\documentclass[pdfusetitle,aps,reprint,prd,twocolumn,superscriptaddress,nofootinbib]{revtex4-1}
\pdfoutput=1

\usepackage{comment}
\usepackage{amssymb}
\usepackage{amsmath}
\usepackage{epsfig}
\usepackage{breakurl}
\usepackage{float}
\usepackage{multirow}
\usepackage{array}
\usepackage{bm}
\usepackage{color}
\usepackage{tikz}
\usepackage[utf8]{inputenc}
\usepackage{braket}
\usepackage{graphicx}
\usepackage{footmisc}
\usepackage{tensor}
\usepackage[normalem]{ulem}
\usepackage{dblfloatfix}
\usepackage[T1]{fontenc}
\usepackage{array}
\usepackage{booktabs}
\usepackage{multirow}
\usepackage{amstext}
\usepackage{subcaption}
\usepackage{multirow}



\usepackage{babel}

\usepackage[force]{feynmp-auto}

\xdefinecolor{mylinkcolor}{rgb}{0,0,0.7}
\usepackage[
	bookmarksnumbered, bookmarksopen, bookmarksopenlevel=2,
	breaklinks=true,
	colorlinks=true, filecolor=mylinkcolor, citecolor=mylinkcolor,
	linkcolor=mylinkcolor, urlcolor=mylinkcolor, menucolor=mylinkcolor,
]{hyperref}
\usepackage{bookmark}

\usepackage{cleveref}
\crefformat{section}{\S#2#1#3} 
\crefformat{subsection}{\S#2#1#3}
\crefformat{subsubsection}{\S#2#1#3}

\usepackage{colortbl}
\definecolor{blue2}{cmyk}{1, 0.1, 0.1, 0}

\definecolor{pyBlue}{RGB}{31, 119, 180}
\definecolor{pyRed}{RGB}{214, 39, 40}
\definecolor{pyGreen}{RGB}{44, 160, 44}
\definecolor{pyBlue2}{RGB}{0, 111, 237}
\definecolor{pyRed2}{RGB}{224, 52, 36}

\definecolor{summersky}{cmyk}{0.71,0.33,0,0.5}
\definecolor{flamingo}{cmyk}{0,0.51,0.71,0.5}
\definecolor{rp}{cmyk}{0.2, 1, 0.6, 0}
\definecolor{pacificblue}{cmyk}{0.95,0.3,0, 0.5}
\definecolor{gray60}{cmyk}{0.4,0.4,0,0.8}

\renewcommand{\vec}[1]{\mathbf{#1}}

\usetikzlibrary{decorations.markings}
\usepackage{cleveref}
\crefformat{section}{\S#2#1#3}
\crefformat{subsection}{\S#2#1#3}
\crefformat{subsubsection}{\S#2#1#3}

\makeatletter
\def\simgt{\mathrel{\lower$\frac{5}{2}$pt\vbox{\lineskip=0pt\baselineskip=0pt
           \hbox{$>$}\hbox{$\sim$}}}}
\def\simlt{\mathrel{\lower$\frac{5}{2}$pt\vbox{\lineskip=0pt\baselineskip=0pt
           \hbox{$<$}\hbox{$\sim$}}}}

\def\tfrac#1#2{{\textstyle \frac{#1}{#2}}}
\makeatother

\def\spa#1.#2{\left\langle#1\,#2\right\rangle}
\def\spb#1.#2{\left[#1\,#2\right]}
\def\sand#1.#2.#3{%
\left\langle#1{\vphantom1}\right|{#2}\left|#3\right]}%
\def\sandmp#1.#2.#3{%
\left\langle#1{\vphantom1}\right|{#2}\left|#3\right]}%
\def\sandpm#1.#2.#3{%
\left[#1{\vphantom1}\right|{#2}\left|#3\right\rangle}%
\def\sandmm#1.#2.#3{%
\left\langle#1{\vphantom1}\right|{#2}\left|#3\right\rangle}%
\def\sandpp#1.#2.#3{%
\left[#1{\vphantom1}\right|{#2}\left|#3\right]}%

\renewcommand{\imath}{\mathrm{i}}

\newcommand{\be}{\begin{equation}}
\newcommand{\ee}{\end{equation}}

\def\S{{\mathbb S}}

\allowdisplaybreaks

\begin{document}

\title{Bootstrapping non-unitary CFTs}
\author{Yu-tin Huang}
\email{yutin@phys.ntu.edu.tw}
\affiliation{Department of Physics and Center for Theoretical Physics, National Taiwan University, Taipei 10617, Taiwan}
\affiliation{Physics Division, National Center for Theoretical Sciences, Taipei 10617, Taiwan}
\affiliation{Max Planck{-}IAS{-}NTU Center for Particle Physics, Cosmology and Geometry, Taipei 10617, Taiwan}
\author{Shao-Cheng Lee}
\email{ken6020255@gmail.com;}
\affiliation{Department of Physics and Center for Theoretical Physics, National Taiwan University, Taipei 10617, Taiwan}
\author{Henry Liao}
\email{henryliao.physics@gmail.com}
\affiliation{Department of Physics and Center for Theoretical Physics, National Taiwan University, Taipei 10617, Taiwan}
\author{Justinas Rumbutis}
\affiliation{Department of Physics and Center for Theoretical Physics, National Taiwan University, Taipei 10617, Taiwan}
\affiliation{Institute for Mathematics, Academia Sinica, Taiwan}

\email{jr3918@as.edu.tw}

\begin{abstract}
We introduce a non-unitary-compatible numerical bootstrap strategy based on the statistical stability of OPE data inferred from crossing at multiple cross-ratios. For a trial spectrum, crossing determines OPE coefficients whose residual cross-ratio dependence directly measures the truncation error. This defines a scalar objective on the space of spectra, allowing bootstrap searches without imposing unitarity. Applied to two-dimensional Virasoro blocks, the method reproduces known A-series minimal models, including non-unitary examples, and yields candidate truncated solutions for 
c>1 with crossing violation comparable to that of minimal models. More generally, our framework provides a practical route to solving bootstrap constraints beyond the convex, unitary setting.

\end{abstract}

\maketitle

\section{Introduction}

Since the seminal work of Rattazzi, Rychkov, Tonni, and Vichi~\cite{Rattazzi:2008pe}, the numerical conformal bootstrap has developed into a powerful framework for constraining conformal field theories (CFTs) across spacetime dimensions~\cite{Polyakov:1974gs,RevModPhys.91.015002}. By combining crossing symmetry with unitarity, one recasts the bootstrap as a convex optimization problem and obtains rigorous bounds on operator dimensions and OPE coefficients. This perspective has led to remarkably precise results, especially when interesting theories lie on the boundary of the allowed region in CFT data space or form isolated islands under additional assumptions, as in the celebrated examples of the 3D Ising and $O(N)$ models~\cite{PhysRevD.86.025022,El-Showk:2014dwa,Kos:2015mba}.

Two complementary challenges nevertheless remain. First, the optimization-based formulation is best suited to delineating the boundary of theory space, whereas it is also desirable to locate theories in its interior. Second, the convex formulation relies crucially on unitarity, while non-unitary CFTs are of broad interest, appearing for example in renormalization-group flows near complex fixed points and in more general holographic settings. This motivates a direct search for solutions to crossing symmetry that does not assume positivity.

This perspective is not new. Gliozzi showed that, given sufficient information about the fusion algebra of low-lying operators, one can detect truncated approximate solutions by analyzing the rank of crossing constraints expanded around the symmetric point~\cite{Gliozzi:2013ysa,Belavin:1984vu}. More recently, non-convex search strategies have been explored using Monte Carlo methods~\cite{Laio:2022ayq} and machine-learning-inspired approaches~\cite{Kantor:2021kbx,K_ntor_2022,Kantor:2022epi}. In practice, however, searching directly over scaling dimensions, spins, and OPE coefficients remains challenging, and successful searches typically require substantial prior input or restriction to neighborhoods of known theories.

In this Letter, we propose a different strategy. For a candidate spectrum, the crossing equations can be inverted to infer the corresponding OPE coefficients. For an exact solution, the inferred OPE data are independent of the choice of cross-ratios. For an approximate or truncated spectrum, by contrast, the inferred OPE coefficients fluctuate as the cross-ratios are varied. The statistical spread of these inferred OPE coefficients therefore provides a direct quantitative measure of proximity to a true solution.\footnote{A related use of the spread of inferred OPE data to exclude unphysical spectra appeared in~\cite{Picco:2016ilr}.} This observation effectively removes the OPE coefficients from the search space and turns the bootstrap problem into the maximization of a single objective function over the spectrum itself.

As a proof of concept, we apply this idea to two-dimensional CFTs with scalar primaries, using Virasoro conformal blocks and the covariance matrix adaptation evolution strategy (CMA-ES)~\cite{hansen2016cma} as the optimizer; implementation details are given in Appendix~\ref{eq: CMAInt}. The method reproduces known A-series minimal models, including non-unitary examples, in both exact and truncated settings. It also yields stable truncated candidate spectra for non-unitary theories with $c>1$ whose crossing violation is comparable to that of minimal models. These results suggest that the statistical stability of inferred OPE data provides a practical route to bootstrap searches beyond the convex, unitary regime.

\section{The objective function}\label{sec:reward}

We consider the four-point function of identical scalar primaries $\phi$ of scaling dimension $\Delta_\phi$,
\begin{equation}
\langle \phi(x_1)\phi(x_2)\phi(x_3)\phi(x_4)\rangle
= \frac{\mathcal{G}(z,\bar z)}{x_{12}^{2\Delta_\phi} x_{34}^{2\Delta_\phi}},
\end{equation}
with cross-ratios defined by $\frac{x_{12}^2x_{34}^2}{x_{13}^2x_{24}^2}=z\bar z$ and $\frac{x_{14}^2x_{23}^2}{x_{13}^2x_{24}^2}=(1-z)(1-\bar z)$. The reduced correlator admits the conformal block expansion~\cite{BELAVIN1984333,Dolan:2000ut}
\begin{equation}\label{eq:sumrule-uv}
\mathcal{G}(z,\bar z)
=
\sum_{\mathcal O}
\big(f_{\phi\phi\mathcal O}\big)^2\,
g_{\Delta_{\mathcal O},s_{\mathcal O}}^{\Delta_\phi}(z,\bar z),
\end{equation}
where $f_{\phi\phi\mathcal O}$ are OPE coefficients and $g_{\Delta,s}^{\Delta_\phi}(z,\bar z)$ are scalar-external conformal blocks. Crossing symmetry may then be written as
\begin{equation}\label{eq:Crossing}
\sum_k \big(f_{\phi\phi k}\big)^2\,G_{\Delta_k,s_k}^{\Delta_\phi}(z,\bar z)
=
\mathrm{vac}(z,\bar z),
\end{equation}
where $G_{\Delta_k,s_k}^{\Delta_\phi}$ denotes the difference between the $s$- and $t$-channel blocks, and $\mathrm{vac}(z,\bar z)$ is the vacuum contribution.

In practice we search over truncated spectra with $N$ exchanged operators, so it is convenient to rewrite crossing as
\begin{equation}\label{crossing_eq_u}
\sum_{k=1}^{N}
\big(f_{\phi\phi k}\big)^2\,G_{\Delta_k,s_k}^{\Delta_\phi}(z,\bar z)
+ e(z,\bar z)
=
\mathrm{vac}(z,\bar z),
\end{equation}
where $e(z,\bar z)$ collects the contribution of omitted operators with $\Delta>\Delta_N$.\footnote{In practice there is also numerical error from the evaluation of Virasoro blocks.} Near the symmetric point $z=\bar z=\tfrac12$, the contribution of high-dimension operators is exponentially suppressed~\cite{Pappadopulo:2012jk}, so for sufficiently well-truncatable spectra the error term can be made small.

We now evaluate Eq.~\eqref{crossing_eq_u} on a set of points $\{(z_j,\bar z_j)\}$ in a neighborhood of the symmetric point. Writing $C_k\equiv \big(f_{\phi\phi k}\big)^2$, $G_{k,j}\equiv G_{\Delta_k,s_k}^{\Delta_\phi}(z_j,\bar z_j)$, and $v_j\equiv \mathrm{vac}(z_j,\bar z_j)$, the crossing equations take the matrix form
\begin{equation}\label{crossing_matrix}
G_{\{z\}}\cdot \vec C + \vec e_{\{z\}}=\vec v_{\{z\}}.
\end{equation}
If we neglect $\vec e_{\{z\}}$, then for each choice of sampled cross-ratios we infer
\begin{equation}
\vec C_{\{z\}} = G_{\{z\}}^{-1}\cdot \vec v_{\{z\}}.
\end{equation}
For an exact solution, $\vec C_{\{z\}}$ is independent of the chosen sample. For a truncated or approximate spectrum, however, the inferred OPE coefficients fluctuate with $\{z\}$. Their statistical spread therefore directly measures the residual crossing error:
\begin{equation}\label{eq:Cinv}
\vec C_{\{z\}}
=
\vec C + G_{\{z\}}^{-1}\cdot \vec e_{\{z\}},
\qquad
\sigma_z(\vec C_{\{z\}})
=
\sigma_z\!\left(G_{\{z\}}^{-1}\cdot \vec e_{\{z\}}\right).
\end{equation}

In practice, the proximity of the sampled points can make $G_{\{z\}}$ ill-conditioned. We therefore take $N_z\gg N$ samples and infer $\vec C_{\{z\}}$ by least squares, which leads to the objective function
\begin{equation}\label{eq:reward}
R
=
-\sum_{i=1}^{N_r}
\log\!\left|
\frac{\sigma(C_{i\{z\}})}{\mathrm{Mean}(C_{i\{z\}})}
\right|,
\end{equation}
where $N_r$ counts the low-lying inferred OPE coefficients included in the reward. Large $R$ therefore signals a spectrum whose inferred OPE data are stable under variations of the sampled cross-ratios, and hence approximately satisfy crossing. When comparing truncations with different numbers of exchanged operators, we usually take $N_r=1$ and focus on the lowest-dimension exchanged state, making the reward directly comparable across truncations.

The key advantage is that the OPE coefficients are removed from the nonlinear search space: instead of scanning over $\{\Delta_k,s_k,(f_{\phi\phi k})^2\}$, we search only over $\{\Delta_k,s_k\}$.

\section{The 2D problem}\label{sec 2d}

We test the construction in two-dimensional CFTs with scalar primaries, using Virasoro conformal blocks. As benchmarks we use A-series minimal models, reviewed briefly in Appendix~\ref{sup: minimal models}. The search variables are the central charge $c$, the external conformal weight $h_{\rm ext}$, and the internal weights $h_{{\rm int},i}$. We begin with a single exchanged primary and enlarge the truncation as needed. Virasoro blocks are evaluated numerically using Zamolodchikov-type recursion relations truncated at finite order.\footnote{We implemented the recursive algorithm of Ref.~\cite{Chen_2017} on GPU and use $40$ recursion steps in the present work.}

We study four search regions, organized by the number of internal states appearing in the minimal models they contain: one, two, three, and five. The search is iterative. If the best reward for a search with $N$ exchanged states stays below a reference threshold, we repeat the search with $N+1$ states; if the threshold is exceeded, we still test larger truncations and declare the spectrum stabilized only when the reward no longer improves and the newly inferred OPE coefficients are statistically consistent with zero.

\begin{figure}[!ht]
    \centering
    \includegraphics[width=0.5\textwidth]{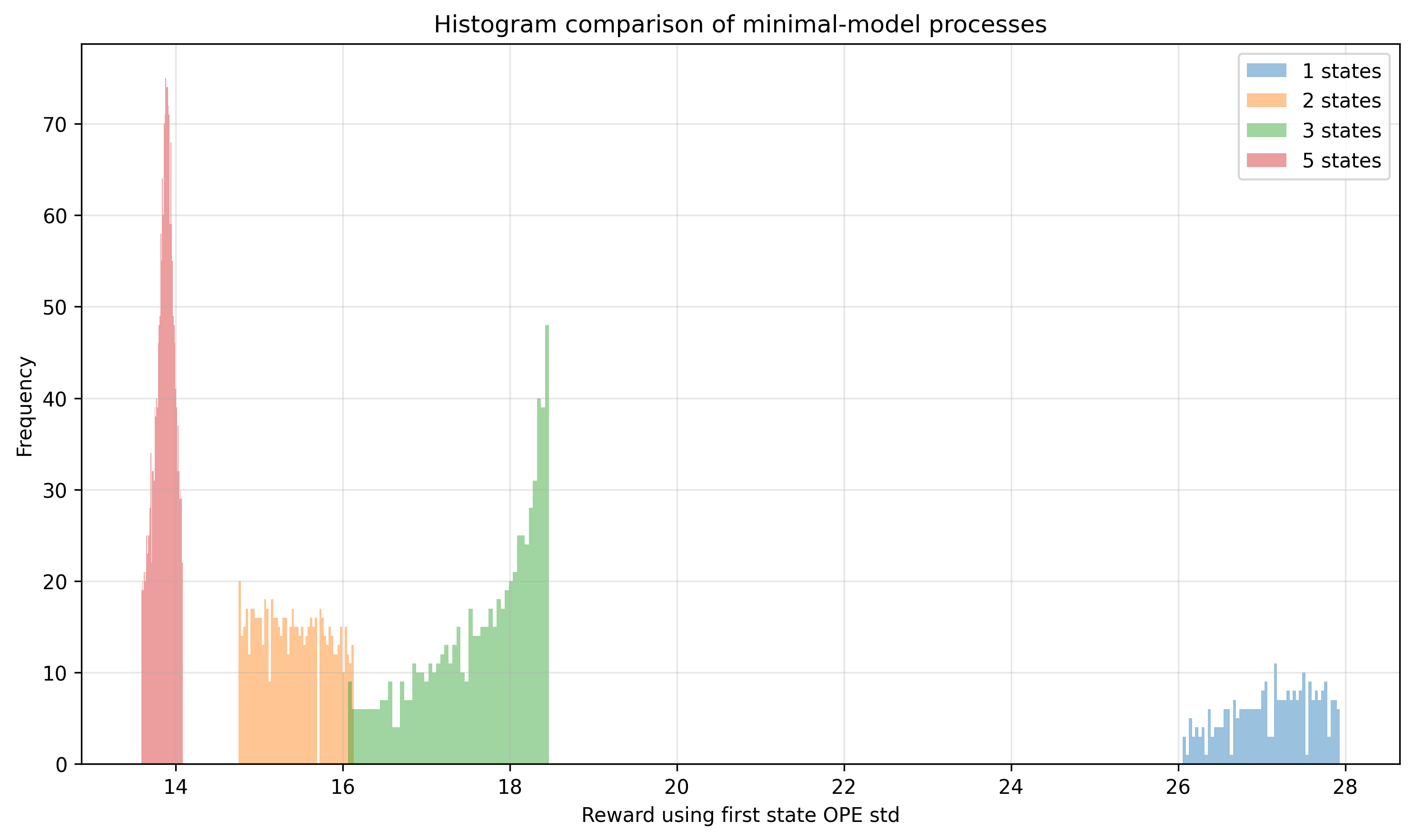}
    \caption{Reward distributions for minimal-model processes in the four search regions. Their $N_r=1$ values motivate the working threshold used below.}
    \label{fig:rew histogram}
\end{figure}

To calibrate the threshold, we evaluate the $N_r=1$ reward for known minimal models in the four search regions. Figure~\ref{fig:rew histogram} suggests the empirical benchmark
\begin{equation}
R_{N_r=1}\simeq 10
\end{equation}
provides a useful empirical criterion for identifying good approximate solutions.

\subsection{Exact spectra for $c<1$}

We first consider regions in which the known solutions involve at most three exchanged Virasoro primaries, so that no physical truncation error is expected. The only residual error then comes from the finite-order evaluation of the Virasoro blocks.

\begin{figure}[!ht]
    \centering
    \includegraphics[width=0.5\textwidth]{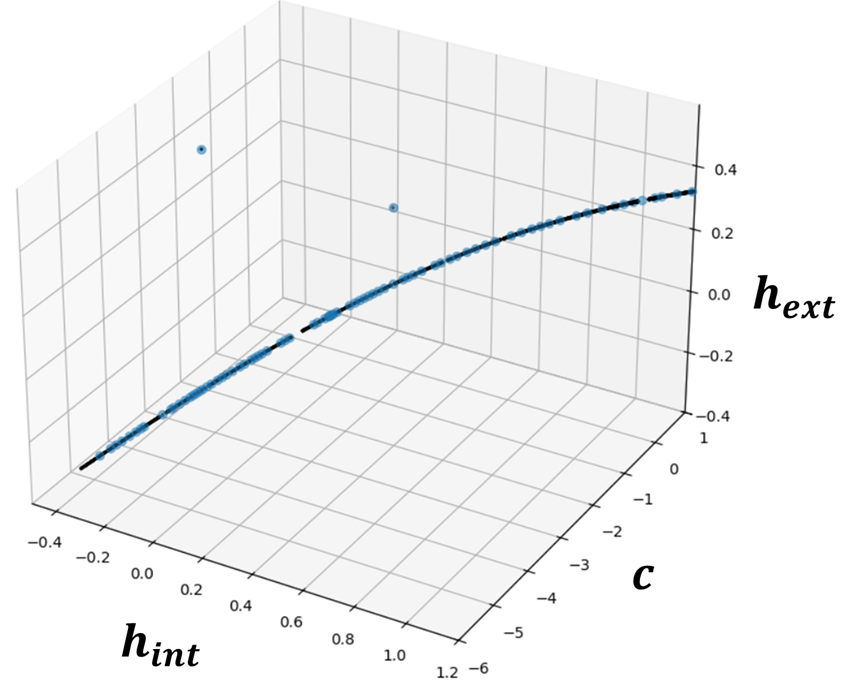}
    \caption{Single-state search in the $c<1$ region. Blue points are search results and black points are analytic minimal-model values.}
    \label{fig:minimal_model_result}
\end{figure}

The first search window,
\begin{equation}
c\in[-6,1],\qquad
h_{\rm ext}\in[-0.4,0.6],\qquad
h_{\rm int}\in[-0.5,1.2],
\end{equation}
contains minimal models with a single exchanged primary. The results, shown in Fig.~\ref{fig:minimal_model_result}, align closely with the analytic minimal-model values.

\begin{figure}[h]
    \centering
    \includegraphics[width=0.5\textwidth]{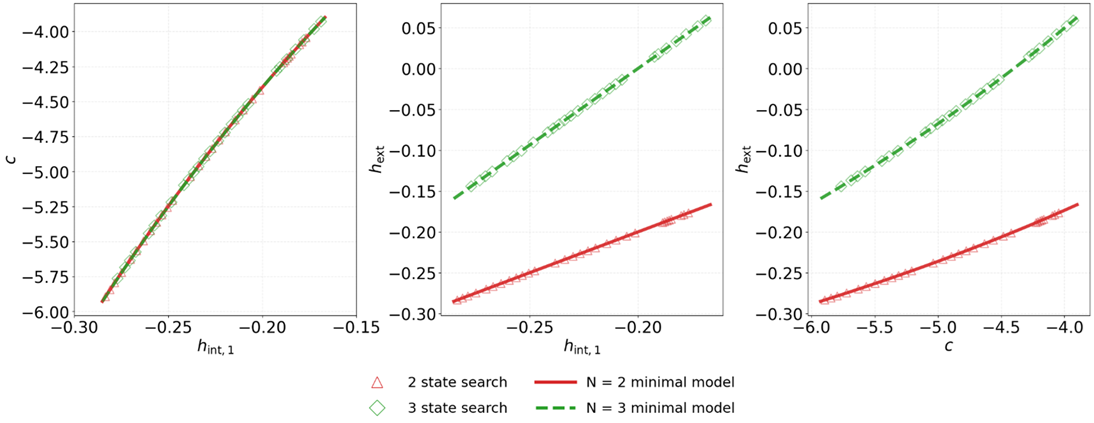}
    \caption{Search results for non-unitary CFTs with $2$ and $3$ internal operators with similar $\left(h_{int,1}, c, h_{ext}\right)$ as Yang-Lee CFT.}
    \label{fig:nonunitary_minimal_model_result_1}
\end{figure}

We next focus on the region surrounding the Yang-Lee theory. Using narrow search windows centered on that region (listed in Appendix~\ref{app:supp_tables}), the resulting two- and three-state scans, shown in Fig.~\ref{fig:nonunitary_minimal_model_result_1}, again agree well with the expected minimal-model spectra; additional projections are collected in Appendix~\ref{sec:further_results}. These examples show that when the exact spectrum lies within the truncation, the reward function reliably identifies the correct CFT data.

\subsection{Truncated spectra for $c<1$}

We now turn to a region in which the relevant minimal models contain five exchanged primaries. Starting from a one-state ansatz, we increase the truncation until the best $N_r=1$ reward exceeds the reference threshold. The search window and the full reward table are given in Appendix~\ref{app:supp_tables}.

For $N<4$, the reward remains below threshold. At $N=4$, the threshold is exceeded, and the resulting spectra already approximate the first four operators of known five-state processes; representative examples are given in Appendix~\ref{sec:selection_of_minimal_models}. At $N=5$, the reward improves substantially. Increasing the truncation to $N=6$ yields no meaningful improvement, while the inferred OPE coefficient of the sixth state is statistically consistent with zero. Since the first five parameters of the best five- and six-state solutions agree at the percent level, we conclude that the search has stabilized at five states.

\begin{figure}[h]
    \centering
    \includegraphics[width=0.5\textwidth]{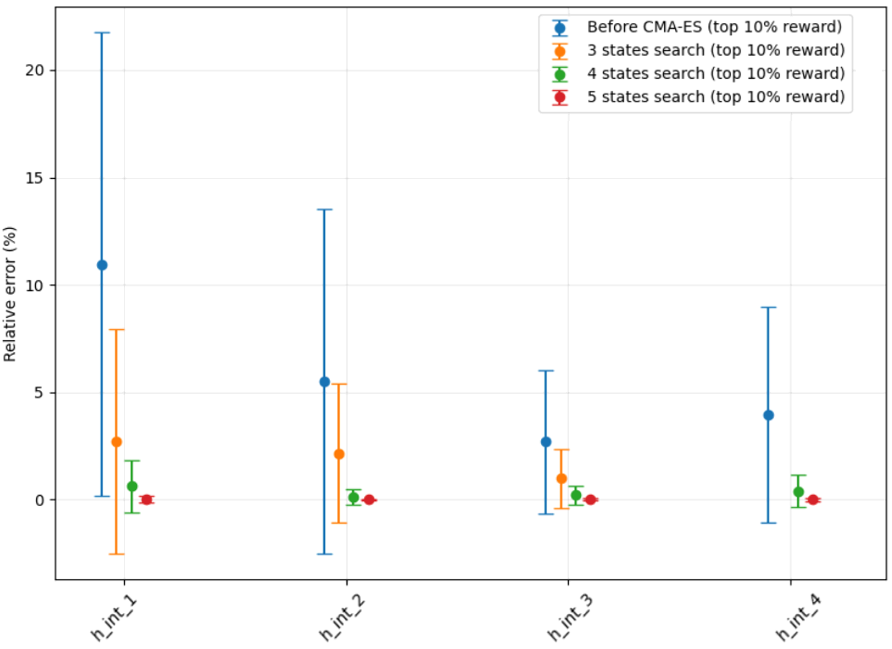}
    \caption{Relative errors for searches in the region containing five-state processes. CMA-ES substantially sharpens the agreement with nearby minimal-model spectra.}
    \label{fig:n=5}
\end{figure}

To quantify agreement with the exact minimal-model spectra in this seven-dimensional search space, we compare each found point with the nearest minimal-model spectrum using the procedure described in Appendix~\ref{sec:error_bar}. The resulting relative-error distributions are shown in Fig.~\ref{fig:n=5}. Relative to the initial random population, CMA-ES sharply improves the agreement with the minimal-model loci, providing a nontrivial validation of the reward function in a genuinely truncated setting.

\section{Candidate non-unitary spectra with $c>1$}

We next apply the method to non-unitary spectra with $c>1$. As before, we begin with a small number of exchanged states and increase the truncation until the $N_r=1$ reward exceeds the reference value established from the minimal-model benchmarks.

\begin{figure}[h]
    \centering
    \includegraphics[width=0.5\textwidth]{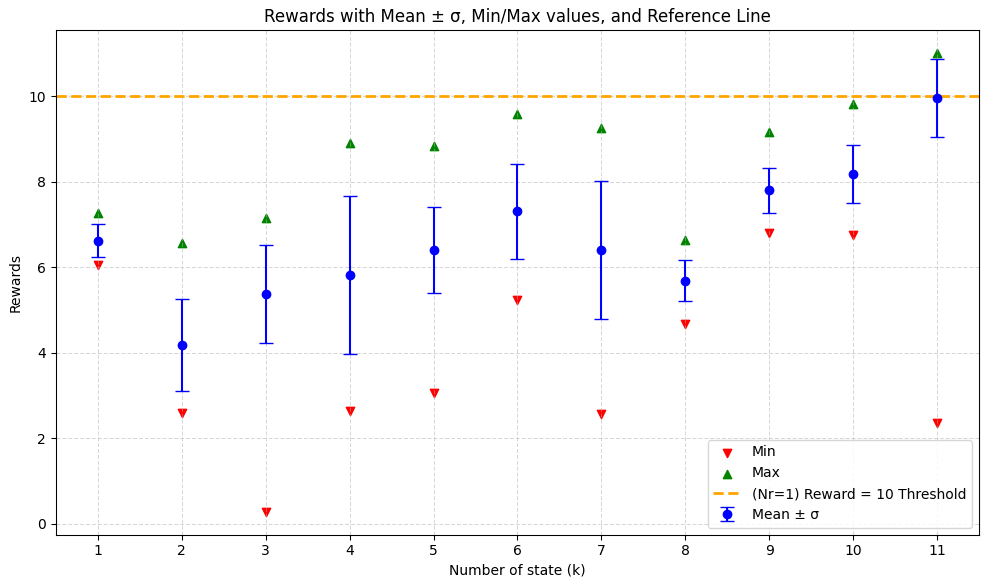}
    \caption{Initial scan over truncation level in the non-unitary $c>1$ search. Spectra with $N_r=1$ reward above threshold first appear at $11$ states.}
    \label{fig:state_scan}
\end{figure}

The initial scan is performed over the broad search window listed in Appendix~\ref{app:supp_tables}. For a $N$-state ansatz, we use the first $N$ entries of the listed ranges for the internal weights together with the corresponding ranges of $c$ and $h_{\rm ext}$. The resulting rewards are shown in Fig.~\ref{fig:state_scan}: spectra with $N_r=1$ reward above threshold first appear at $N=11$.

To test stability under larger truncations, we select a candidate from the $11$-state scan and perform a refined search in a narrower window, again detailed in Appendix~\ref{app:supp_tables}, now allowing for $12$ and $13$ exchanged states. The average $N_r=1$ reward increases from $9.86$ at $N=11$ to $10.55$ at $N=12$ and then remains essentially unchanged, $10.53$, at $N=13$, suggesting that the low-lying spectrum has stabilized.

This conclusion is supported by the inferred OPE data. For the highest-reward $13$-state solution (see Table \ref{table_ope} in Appendix~\ref{app:supp_tables}), the inferred OPE coefficients of the $12$th and $13$th states are about three orders of magnitude smaller than that of the $11$th state. Moreover,
\begin{equation}
\left|\frac{\sigma(C_{12})}{\mathrm{Mean}(C_{12})}\right|,\qquad
\left|\frac{\sigma(C_{13})}{\mathrm{Mean}(C_{13})}\right|,
\end{equation}
are much larger than the corresponding quantity for the $11$th state, indicating that the extra coefficients are not statistically significant. Since heavier conformal blocks are exponentially suppressed near the symmetric point~\cite{Pappadopulo:2012jk}, their contribution to the sampled crossing equations is negligible.

\begin{figure}[h]
    \centering
    \includegraphics[width=0.5\textwidth]{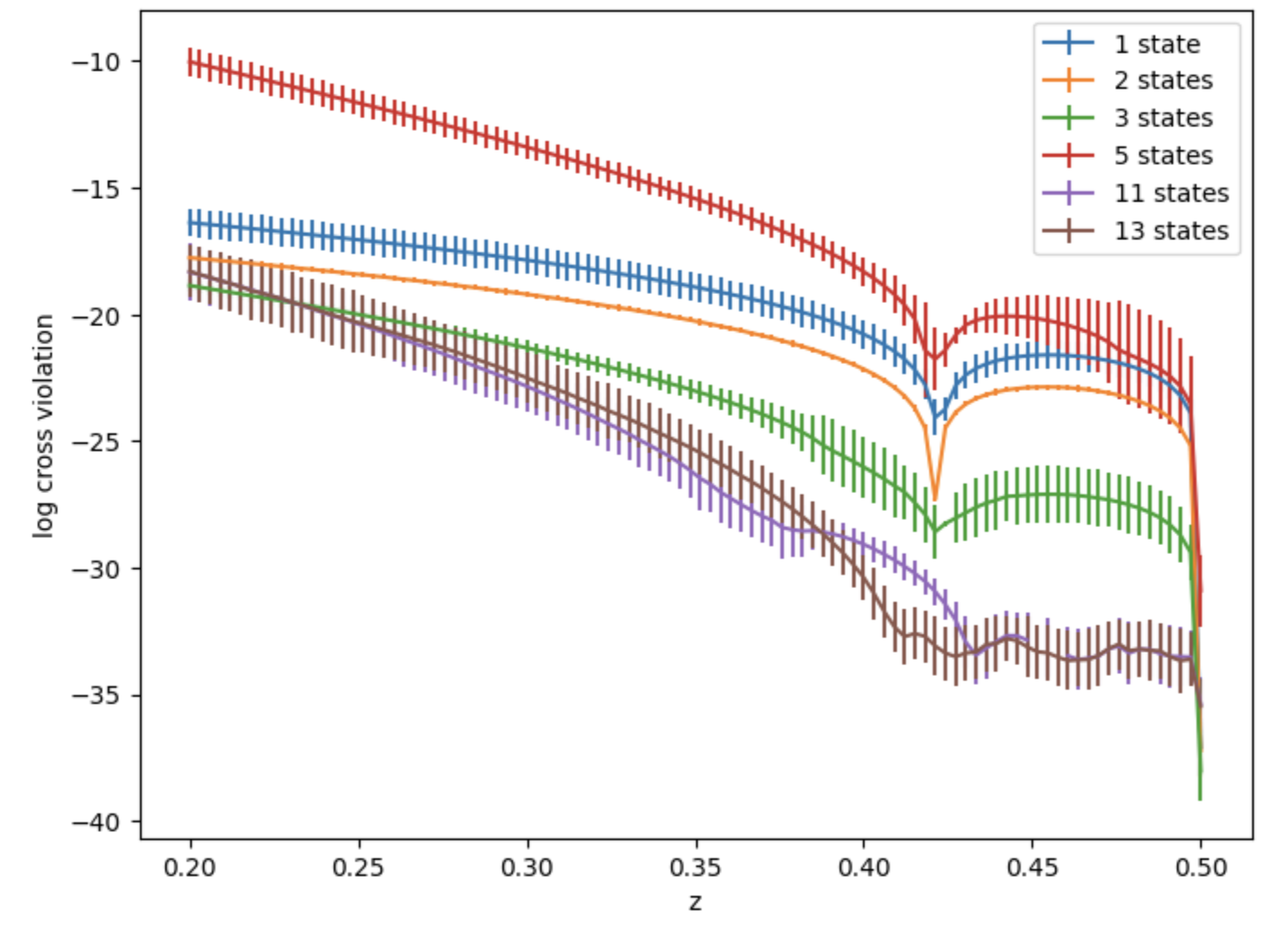}
\caption{Crossing violation at $\bar z=1/2$ for minimal-model benchmarks and for the candidate $11$- and $13$-state $c>1$ spectra. Solid lines denote means and error bars denote standard deviations across spectra with the same truncation.}
    \label{fig:cross viol}
\end{figure}

A complementary diagnostic is the crossing violation itself. Figure~\ref{fig:cross viol} compares the candidate $11$- and $13$-state spectra with minimal models containing $1$, $2$, $3$, and $5$ exchanged states. Across a broad range of $z$, the candidate $c>1$ spectra have crossing violation comparable to the minimal-model benchmarks, and the similarity of the $11$- and $13$-state curves again indicates that the truncation has stabilized by $11$ states.

Taken together, these observations provide evidence for stable truncated spectra consistent with non-unitary solutions of crossing symmetry at $c>1$.  Additional examples are provided in~\cite{githubrepo:nucftb}, and the five highest-reward $13$-state solutions are listed in Table~\ref{tab:13_st_examples} of Appendix~\ref{sec:further_results}.

\section{Conclusion and outlook}

In this work we propose a simple objective function for searching CFT spectra satisfying the crossing equation, that is based on the statistics of solutions of OPE coefficients across many different $z, \bar{z}$ points. This allows us to study solutions to the bootstrap equations for non-unitary CFTs, where conventional convex optimization approaches are no longer applicable. We implemented this framework in two-dimensional CFTs with Virasoro blocks using a fully GPU-parallelized search. In regions where the exact spectrum lies within the truncation, the method reproduces the spectra of A-series minimal models, including non-unitary examples. In regions where truncation is unavoidable, it identifies stable approximate spectra and correctly determines when additional states cease to improve the solution. Extending the search to $c>1$, we found stable truncated candidate spectra whose crossing violation is numerically comparable to that of minimal models.

Several directions deserve further study. Extension to include spin quantum numbers should be straight forward. It will be important to develop a reliable procedure to further test consistency of the spectrum and reconstruct the low lying spectrum. This include extending to mixed correlators and the modularity of torus partition functions~\cite{Benjamin:2026lbj}.  

It would also be interesting to improve the treatment of the truncation error $e(z,\bar z)$ by incorporating analytic expectations for the asymptotic spectrum. In dimensions $d>2$, for example, the light-cone bootstrap implies the existence of large-spin double-twist operators~\cite{Simmons-Duffin:2016wlq, Kusuki:2018wpa}, suggesting a natural way to constrain the tail of the spectrum that is otherwise absorbed into $e$. Since the resulting objective function is structured but generally non-differentiable, it may also provide a useful arena for optimization strategies beyond CMA-ES, including learning-based approaches.

\section*{Acknowledgments}  
We would like to thank Po-Chung Chen, Li-Yuan Chiang, Miranda Cheng, Chong-Sun Chu, David Poland and Ning Su for enlightening discussions. We would especially like to thank Sylvain Ribault for bringing to our attention previous work utilizing the statistical spread of those inferred OPEs.   Y-t H thanks Riken iTHEMS and the Yukawa Institute for
Theoretical Physics at Kyoto University. Discussions during
“Progress of Theoretical Bootstrap” were useful in completing this work. Y-tH is supported by the Taiwan National Science and Technology Council grant 112-2628-M-002-003-MY3 and 114-2923-M-002-011-MY5. S-C Lee is supported by 112-2628-M-002-003-MY3.  H Liao is supported in part by the Ministry of Science and Technology grant 112-2112-M-002-024-MY3 and 112-2628-M-002-003-MY3.

\clearpage
\onecolumngrid
\appendix

\section{Supplemental tables}\label{app:supp_tables}

For convenience, we collect here the search windows and representative spectra referred to in the main text.

\begin{table}[htb]
\centering

\begin{minipage}[t]{0.32\columnwidth}
\centering
\captionof{table}{Search window for the $2$-state Yang-Lee-region scan.}
\label{table2}
\vspace{0.5em}
\begin{tabular}{|c|c|}
\hline
Variable & Search bounds \\
\hline
$h_{int,1}$ & $[-0.3,-0.1]$ \\
$h_{int,2}$ & $[0.15,0.5]$ \\
$c$         & $[-6.5,-4]$ \\
$h_{ext}$   & $[-0.3,-0.2]$ \\
\hline
\end{tabular}
\end{minipage}
\hfill
\begin{minipage}[t]{0.32\columnwidth}
\centering
\captionof{table}{Search window for the $3$-state Yang-Lee-region scan.}
\label{table3}
\vspace{0.5em}
\begin{tabular}{|c|c|}
\hline
Variable & Search bounds \\
\hline
$h_{int,1}$ & $[-0.3,-0.1]$ \\
$h_{int,2}$ & $[0.1,0.5]$ \\
$h_{int,3}$ & $[1.1,2]$ \\
$c$         & $[-7,-3]$ \\
$h_{ext}$   & $[-0.3,0.1]$ \\
\hline
\end{tabular}
\end{minipage}
\hfill
\begin{minipage}[t]{0.32\columnwidth}
\centering
\captionof{table}{Search window for the five-state truncated search.}
\label{table5}
\vspace{0.5em}
\begin{tabular}{|c|c|}
\hline
Variable & Search bounds \\
\hline
$h_{int,1}$ & $[0.23,0.3]$ \\
$h_{int,2}$ & $[0.35,0.4]$ \\
$h_{int,3}$ & $[1.8,2]$ \\
$h_{int,4}$ & $[2.8,3.2]$ \\
$h_{int,5}$ & $[6.3,6.9]$ \\
$c$         & $[0.1,0.23]$ \\
$h_{ext}$   & $[0.85,0.95]$ \\
\hline
\end{tabular}
\end{minipage}

\end{table}

\begin{table}[!h]
    \begin{center}
    \caption{Maximum rewards of the searches with different numbers of states in the region given in Table~\ref{table5}. For the best six-state spectrum, the mean OPE coefficient of the last state is of the same order as its standard deviation, indicating that five states are sufficient to complete the spectrum.}
    \label{table nst}
 \begin{tabular}{ |c|c|c| }
    \hline
    Number of states & Max $N_r=1$ reward & $\left| \frac{\sigma (C_{\rm last})}{\mathrm{Mean}(C_{\rm last})} \right|$ \\
    \hline
    1 & 4.7894  & $4.4651\times 10^{-10}$ \\
    2 & 5.3099  & $5.2581\times 10^{-10}$ \\
    3 & 4.2433  & $3.2996\times 10^{-10}$ \\
    4 & 12.6996 & $1.6291\times 10^{-9}$  \\
    5 & 21.5746 & $6.4221\times 10^{-4}$  \\
    6 & 21.5296 & $6.5360\times 10^{-1}$  \\
    \hline
\end{tabular}

    \end{center}
\end{table}

\begin{table}[htb]
  \centering
  \begin{minipage}[t]{0.48\columnwidth}
    \centering
    \captionof{table}{Initial search window for the non-unitary $c>1$ scan.}
    \label{table5a}
    \vspace{0.5em}
    \begin{tabular}{|c|c|}
      \hline
      Variable & Search bounds \\
      \hline
      $h_{int,1}$ & $[-0.5,-0.3]$ \\
      $h_{int,2}$ & $[0.1,0.3]$ \\
      $h_{int,3}$ & $[0.3,0.5]$ \\
      $h_{int,4}$ & $[0.5,0.7]$ \\
      $h_{int,5}$ & $[0.7,0.9]$ \\
      $h_{int,6}$ & $[0.9,1.1]$ \\
      $h_{int,7}$ & $[1.1,1.3]$ \\
      $h_{int,8}$ & $[1.3,1.5]$ \\
      $h_{int,9}$ & $[1.5,1.7]$ \\
      $h_{int,10}$ & $[1.7,1.9]$ \\
      $h_{int,11}$ & $[1.9,2.1]$ \\
      $c$         & $[1.05,1.25]$ \\
      $h_{ext}$   & $[-0.25,-0.05]$ \\
      \hline
    \end{tabular}
  \end{minipage}
  \hfill
  \begin{minipage}[t]{0.48\columnwidth}
    \centering
    \captionof{table}{Refined search window for the $12$- and $13$-state scans.}
    \label{table6}
    \vspace{0.5em}
    \begin{tabular}{|c|c|}
      \hline
      Variable & Search bounds \\
      \hline
      $h_{int,1}$ & $[-0.4,-0.3]$ \\
      $h_{int,2}$ & $[0.25,0.35]$ \\
      $h_{int,3}$ & $[0.4,0.5]$ \\
      $h_{int,4}$ & $[0.55,0.65]$ \\
      $h_{int,5}$ & $[0.75,0.85]$ \\
      $h_{int,6}$ & $[0.95,1.05]$ \\
      $h_{int,7}$ & $[1.15,1.25]$ \\
      $h_{int,8}$ & $[1.3,1.4]$ \\
      $h_{int,9}$ & $[1.55,1.65]$ \\
      $h_{int,10}$ & $[1.75,1.85]$ \\
      $h_{int,11}$ & $[2.0,2.1]$ \\
      $h_{int,12}$ & $[2.1,3.5]$ \\
      $h_{int,13}$ & $[2.9,4.5]$ \\
      $c$         & $[1.01,1.11]$ \\
      $h_{ext}$   & $[-0.17,-0.07]$ \\
      \hline
    \end{tabular}
  \end{minipage}
\end{table}

\begin{table}[htb]
  \centering
  \begin{minipage}[t]{0.48\columnwidth}
    \centering
    \captionof{table}{Spectrum and inferred OPE coefficients for the highest-reward $k=13$ solution, with $(c,h_{ext})=(1.07662761,-0.11844774)$.}
    \label{table_ope}
    \vspace{0.5em}
    \begin{tabular}{|c|c|c|}
      \hline
      $j$ & $h_{int,j}$ & $C_j$   \\
      \hline
      1&$-0.30347016$ & $1.02405595$ \\
      2&$0.33772710$ & $-0.11433376$ \\
      3&$0.48003697$ & $2.720435150$ \\
      4&$0.56931245$ & $-6.10600342$ \\
      5&$0.82098216$ & $12.38668966$ \\
      6&$1.03021669$ & $-21.7457633$ \\
      7&$1.22721684$ & $25.75052912$ \\
      8&$1.37006652$ & $-15.6260878$ \\
      9&$1.60408115$ & $3.273236841$ \\
      10&$1.82665527$ & $-0.6093234$ \\
      11&$2.04846001$ & $0.07390606$ \\
      12&$2.84190083$ & $-0.00008038$ \\
      13&$3.21548438$ & $-0.0000190$ \\
      \hline
    \end{tabular}
  \end{minipage}
\end{table}
\newpage
\section{Specs of search}\label{eq: CMAInt}

Covariance Matrix Adaptation Evolution Strategy (CMA-ES) is a stochastic,
derivative-free algorithm for continuous, noisy black-box optimization. 
It works by evolving an initial random distribution of points toward regions of better rewards.

More concretely, CMA-ES maintains a multivariate normal search distribution
\begin{align}
    x \sim \mathcal{N}(\mu, \sigma^2 \mathcal{C}),
\end{align}
where $\mu$ is the mean (current best guess), $\sigma$ is a global step size, and $\mathcal{C}$ is a positive-definite covariance matrix. 
At each round (or epoch), several candidate solutions are drawn from this distribution, say $N_c$ of them.
They are then ranked according to their rewards from top to bottom.
The mean $\mu$ is updated to a weighted average of the best candidates, the step size $\sigma$ is updated depending on whether progress appears steady, and the covariance $\mathcal{C}$ is reshaped so that it stretches along directions that repeatedly led to good solutions and squeezes along directions that performed poorly. 
This loop repeats until $\sigma$ becomes very small or the improvements stall according to some stopping criterion.

The complete setup includes two parts: CMA-ES and the evaluation of the reward, that is,~\eqref{eq:reward}.
However, there are two difficulties in evaluating this reward.
First, we need to take large enough $N_z$ to reduce the statistical noise.
Second, multiple evaluations of the Virasoro block is required and thus is expensive.
Hence, we evaluate the reward on the GPU.
Also, to reduce the overhead due to data transfer between GPU memory to RAM, we choose to implement CMA-ES on GPU.

To put CMA-ES on the GPU, we adopt the solver from \textit{Evotorch}.
In particular, we use \textit{evotorch.algorithms.CMAES}, which wraps the one in \textit{pycma} so that it can be executed on the GPU backend.
For the evaluation of the reward, we can parallelize the computation across $N_c$, $N_z$ and $N$ internal states.
That is, during each epoch, we have $N_c$ candidates, each of which has $N_z\times N$ components in $G_{\{z\}}$ in \eqref{crossing_matrix}, and we calculate these $N_c\times N_z\times N$ terms with a single pass through the GPU.
Hence, we can significantly reduce the runtime.

Since CMA-ES is a self-adaptive method, it is almost parameter-free once we set the initial conditions (to start the search) and stopping condition (to prevent time waste).
In fact, we only need to choose (1) an initial step size $\sigma_0$ reflecting the scale of the search space, (2) the population size $N_p$, (3) the maximum number of steps for evolution (also known as generation) $N_g$, and (4) the tolerance $t_\sigma$ to accept a solution when $\sigma$ is below this value.
The initial mean $\mu_0$ is random sampled from the search space while the covariance matrix $\mathcal{C}$ is initialized to identity.
The code and the setting we use to run the search are provided in \cite{githubrepo:nucftb}.

In our case, we run the code \cite{githubrepo:nucftb} on a single Nvidia A5500.
As mentioned previously, for each epoch, the runtime is only a single pass on the GPU as long as there are enough threads for the computation.
In our experience, GPU threads on A5500 are not exhausted when we calculate with up to $(N_c,N_z,N)=(500, 400,5)$.
In such case, with the level of recursion for Virasoro block evaluation being $40$, the runtime is about $2$ hours to search for $1000$ points.
As long as we have sufficient number of GPU threads, the runtime does not scale with $N_c$, $N_z$ and $N$, and scale linearly with the recursion level and number of points to search.


\section{2d Minimal Model CFT}\label{sup: minimal models}
In two-dimensional CFT, local primary operators are labeled by holomorphic and anti-holomorphic weights $(h,\bar h)$, with scaling dimension and spin
\begin{equation}
\Delta = h+\bar h, \qquad s = h-\bar h \in \mathbb{Z}.
\end{equation}
The Virasoro minimal models $M(p,q)$ are labeled by coprime integers $p>q\ge 2$, with central charge
\begin{equation}
c = 1 - \frac{6(p-q)^2}{pq}.
\end{equation}
Primary fields are indexed by Kac labels $(r,s)$ with $1 \le r \le q-1$ and $1 \le s \le p-1$, modulo the identification
\begin{equation}
(r,s) \sim (q-r,\, p-s).
\end{equation}
Their conformal weights are
\begin{equation}
h_{r,s} = \frac{(pr - qs)^2 - (p-q)^2}{4pq}.
\end{equation}
In the A-series models $\bar h = h$, so that
\begin{equation}
\Delta = 2h, \qquad s = 0 ,
\end{equation}
while in non-diagonal invariants the left and right Kac labels may differ, giving nonzero spin. \\
A special property of Virasoro minimal models is that their spectrum contains only finitely many Virasoro primary operators. 
Therefore, in principle using exact Virasoro conformal blocks we could solve  \eqref{eq: Crossing} exactly. In practice, however, there is still error due to a numerical evaluation of the Virasoro blocks using finite recursion iterations. \\

Following \cite{Chen_2017} we calculate the Virasoro blocks

\begin{equation}
\mathcal{V}(z,c, h_{\rm ext}, h, N) = 
(16q)^{\,h - \frac{c-1}{24}}
\;z^{\frac{c-1}{24} - 2\,h_{\rm ext}}
\;(1-z)^{\frac{c-1}{24} - 2\,h_{\rm ext}}
\;\theta_3(q)^{\frac{c-1}{2} - 16\,h_{\rm ext}}
\;\sum_{j=0}^{\lfloor N/2\rfloor} q^{2j}\;\mathrm{H}\bigl(c, h_{\rm ext}, h_{\rm ext}, h, N\bigr),
\end{equation}
as truncated series in 
\begin{equation}
q(z) \;=\; \exp\!\Big[-\pi\,\frac{K(1-z)}{K(z)}\Big], 
\end{equation}
where $K$ is the complete elliptic integral of the first kind:
\begin{equation}
K(k) \;=\; \int_{0}^{\frac{\pi}{2}} \frac{d\varphi}{\sqrt{1 - k^2 \sin^2\varphi}}.
\end{equation}
$g_{\Delta_k,s_k}^{\Delta_\phi}(z,\bar z)$ in \eqref{eq:sumrule-uv} can be expressed as
\begin{equation}
g_{\Delta_k,s_k}^{\Delta_\phi}(z,\bar z)=\mathcal{V}(z,c, h_{\rm ext}, h_k, N)\mathcal{V}(\bar{z},c, h_{\rm ext}, h_k, N),
\end{equation}
therefore our non-linear search variables are $c$, $h_{\rm ext}$ and $\{h_k\}$.

\section{Further results}\label{sec:further_results}

In this section, we collect additional results discussed in the main text. Firstly, we show projections of results for 2 and 3 state searches around Yang-Lee model. Each projection is compared with known minimal models, which are aligns well.

\begin{figure}[h]
    \centering
    \includegraphics[width=0.8\textwidth]{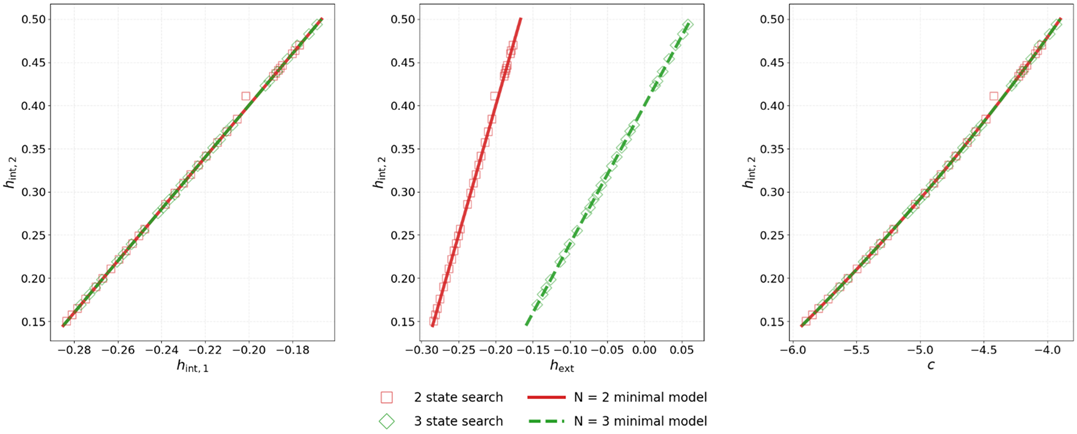}
    \caption{Search results around Yang-Lee CFT region with $2$ and $3$ internal operators in projection to ($O$, $h_{int,2}$) where $O\in\{h_{int,1}, h_{ext}, c\}$.
    The line of minimal model represents minimal models with $p,q\leq 200$.}
    \label{fig:nonunitary_minimal_model_result_2}
\end{figure}

\begin{figure}[h]
    \centering
    \includegraphics[width=0.8\textwidth]{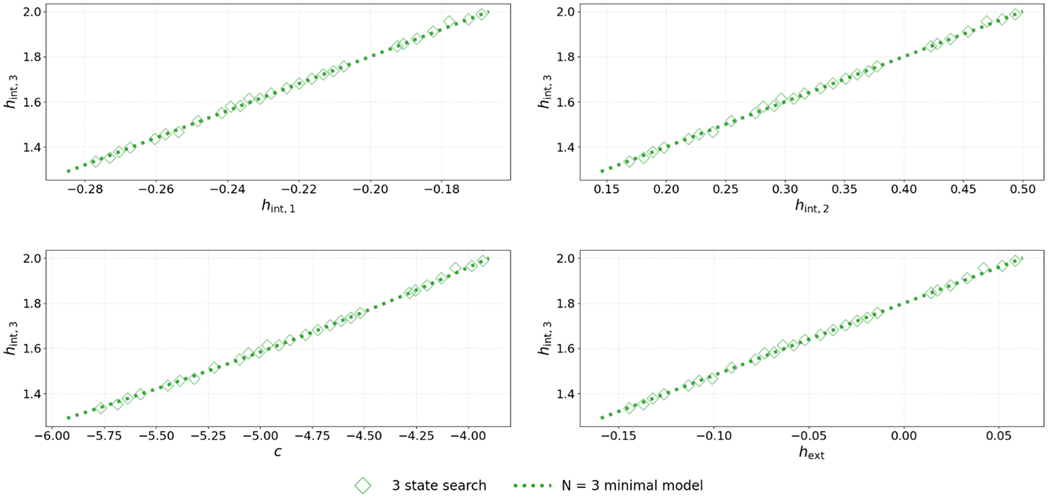}
    \caption{Search results around Yang-Lee CFT region with $3$ internal operators in projection to ($O$, $h_{int,3}$) where $O\in\{h_{int,1}, h_{int,2}, h_{ext}, c\}$.
    The line of minimal model represents minimal models with $p,q\leq 200$.}
    \label{fig:nonunitary_minimal_model_result_3}
\end{figure}


Here we present the top $5$ spectrum with $13$ states in terms of rewards. 
The resulting crossing violation is displayed in Fig \ref{fig:cross viol}. 
\begin{table}[h]
\centering
\caption{Examples of found 13 state spectra with $N_r=1$ reward greater than 10}
\label{tab:13_st_examples}
\begin{tabular}{rcccccccccccc@{\hspace{2em}}ccc}
\toprule
$h_{\text{int}, 1}$ & $h_{\text{int}, 2}$ & $h_{\text{int}, 3}$ & $h_{\text{int}, 4}$ & $h_{\text{int}, 5}$ & $h_{\text{int}, 6}$ & $h_{\text{int}, 7}$ & $h_{\text{int}, 8}$ & $h_{\text{int}, 9}$ & $h_{\text{int}, 10}$ & $h_{\text{int}, 11}$ & $h_{\text{int}, 12}$ & $h_{\text{int}, 13}$ & $c$ & $h_{\text{ext}}$ & $N_r=1$ reward \\ 
\midrule
-0.303 & 0.338 & 0.480 & 0.569 & 0.821 & 1.030 & 1.227 & 1.370 & 1.604 & 1.827 & 2.048 & 2.842 & 3.215 & 1.077 & -0.118 & 12.516 \\
-0.301 & 0.315 & 0.476 & 0.593 & 0.792 & 1.007 & 1.210 & 1.358 & 1.614 & 1.802 & 2.061 & 2.872 & 3.621 & 1.071 & -0.118 & 12.411 \\
-0.300 & 0.349 & 0.472 & 0.563 & 0.760 & 0.992 & 1.213 & 1.375 & 1.616 & 1.792 & 2.069 & 2.855 & 3.881 & 1.078 & -0.118 & 12.331 \\
-0.303 & 0.344 & 0.471 & 0.608 & 0.778 & 1.028 & 1.177 & 1.364 & 1.614 & 1.789 & 2.019 & 2.879 & 3.410 & 1.079 & -0.118 & 12.328 \\
-0.306 & 0.343 & 0.469 & 0.585 & 0.785 & 0.972 & 1.208 & 1.357 & 1.622 & 1.802 & 2.094 & 2.833 & 4.093 & 1.078 & -0.120 & 12.134 \\
\bottomrule
\end{tabular}
\end{table}

\section{Selection of minimal models and comparison with search results}\label{sec:selection_of_minimal_models}

In order to check the robustness of our method, we need to have many minimal models' spectrum as Fig.~\ref{fig:minimal_model_result} shown. 

Minimal model fusion rule
\begin{equation}
\phi_{r_1, s_1} \times \phi_{r_2, s_2} = \sum^{min(r_1+r_2-1, 2q-r_1-r_2-1)}_{r=\left|r_1-r_2\right|+1}\sum^{min(s_1+s_2-1, 2p-s_1-s_2-1)}_{s=\left|s_1-s_2\right|+1} \phi_{r, s}
\end{equation}
By fusion rule, we then have spectrum for any given $(p, q, r, s)$. 
The procedure of the minimal model spectrum computation will be the following.
(1) Set a max $p$.
(2) Scan through all possible $(p, q)$ $\rightarrow$ Compute central charge.
(3) Scan through all possible $(r, s)$ $\rightarrow$ Compute external operator's conformal weight.
(4) Get all internal operators' conformal weight by fusion rule.


To illustrate the result of the selection process, we present five representative examples of identified Nearest Minimal model spectra, including the corresponding $(p, q, r, s)$ parameters, Found spectra, and the relative errors. The following Table \ref{tab:nearest_mm_found_spec_no_hint4} represents the results obtained by truncating a five internal states process to four internal states in the region of Table \ref{table5}, where known five states models are available.
\begin{table}[h!]
\centering
\small
\caption{Nearest Minimal Model (MM) vs.\ Found Spectrum}
\setlength{\tabcolsep}{4pt}
\begin{tabular}{c c ccccccc}
\hline
\textbf{ID} & \textbf{Spectrum} 
 & $h_{\text{int},0}$ & $h_{\text{int},1}$ & $h_{\text{int},2}$ & $h_{\text{int},3}$ & $c$ & $h_{\text{ext}}$ \\
\hline
\multirow{3}{*}{1} & Nearest MM: (118, 81, 3, 2) 
 & 0.286462 & 0.372881 & 1.913580 & 3.113622 & 0.140625 & 0.928411 \\
 & Found 
 & 0.287020 & 0.372803 & 1.914932 & 3.116666 & 0.140634 & 0.929349 \\
 & Relative error 
 & 0.19\% & -0.020\% & 0.070\% & 0.097\% & -0.0064\% & -0.10\% \\
\hline
\multirow{3}{*}{2} & Nearest MM: (427, 300, 3, 2)
 & 0.251819 & 0.405152 & 1.846667 & 2.945152 & 0.244553 & 0.873599 \\
 & Found 
 & 0.252044 & 0.405123 & 1.847260 & 2.947820 & 0.244576 & 0.873992 \\
 & Relative error 
 & 0.089\% & -0.0071\% & 0.032\% & 0.090\% & -0.0094\% & -0.044\% \\
\hline
\multirow{3}{*}{3} & Nearest MM: (330, 227, 3, 2) 
 & 0.283247 & 0.375758 & 1.907489 & 3.098225 & 0.150270 & 0.923398 \\
 & Found 
 & 0.283745 & 0.375704 & 1.908713 & 3.101137 & 0.150278 & 0.924243 \\
 & Relative error 
 & 0.17\% & -0.014\% & 0.064\% & 0.093\% & -0.0053\% & -0.091\% \\
\hline
\multirow{3}{*}{4} & Nearest MM: (482, 337, 3, 2) 
 & 0.258874 & 0.398340 & 1.860534 & 2.979943 & 0.223387 & 0.884912 \\
 & Found 
 & 0.259215 & 0.398285 & 1.861406 & 2.983111 & 0.223426 & 0.885496 \\
 & Relative error 
 & 0.13\% & -0.013\% & 0.046\% & 0.10\% & -0.017\% & -0.066\% \\
\hline
\multirow{3}{*}{5} & Nearest MM: (223, 154, 3, 2) 
 & 0.277270 & 0.381166 & 1.896104 & 3.069478 & 0.168201 & 0.914041 \\
 & Found 
 & 0.277816 & 0.381065 & 1.897435 & 3.072840 & 0.168233 & 0.914957 \\
 & Relative error 
 & 0.19\% & -0.026\% & 0.070\% & 0.10\% & -0.019\% & -0.10\% \\
\hline
\end{tabular}
\label{tab:nearest_mm_found_spec_no_hint4}
\end{table}
In Table \ref{tab:five_six_state_spectra}, we show a six state search in a five state region, neglecting the final state which has near zero OPE coefficient, we find that it recovers the five state solution.  
\begin{table}[h!]
\centering
\small
\caption{Comparison of minimal model spectrum vs.\ the best reward five-state and six-state searches, including relative errors. The predicted OPE coefficient of the last state in the six state search is $\text{OPE} \pm \sigma = -1.203\times 10^{-8} \pm 6.974\times 10^{-9}$.}
\setlength{\tabcolsep}{4pt}
\begin{tabular}{c cccccccc}
\hline
\textbf{Spectrum} 
 & $h_{\text{int},0}$ & $h_{\text{int},1}$ & $h_{\text{int},2}$ & $h_{\text{int},3}$ 
 & $h_{\text{int},4}$ & $h_{\text{int},5}$ & $c$ & $h_{\text{ext}}$ \\
\hline
Five-state MM 
 & 0.2892 & 0.3704 & 1.9187 & 3.1267 & 6.7562 & --     & 0.1324 & 0.9327 \\
Five-state Found 
 & 0.2892 & 0.3705 & 1.9187 & 3.1266 & 6.7568 & --     & 0.1325 & 0.9326 \\
Relative error 
 & -0.0107\% & 0.0074\% & -0.0031\% 
 & -0.0047\% & 0.0087\% & -- 
 & 0.0629\% & -0.0052\% \\
\hline
Six-state Found 
 & 0.2873 & 0.3721 & 1.9151 & 3.1175 & 6.7440 & 7.5403 & 0.1382 & 0.9297 \\
Relative error 
 & -0.66\% & 0.45\% & -0.19\%
 & -0.29\% & -0.18\% & -- & 4.37\%
 & -0.32 \% \\
\hline
\end{tabular}
\label{tab:five_six_state_spectra}
\end{table}

\section{Procedure of error analysis}\label{sec:error_bar}

As mentioned in Sec.~\ref{sec:reward}, we are focusing on the crossing symmetry of four-point functions with identical scalars.
After decomposing it using conformal blocks, we parametrize our search space into $\left(\vec{h}_{int}, h_{ext}, c\right)$, in which we can plot known theories, such as minimal models, to qualify our search results.
When theories are plotted in this space, they in general clustered into multiple lines in the space, see any figures in the main text.
That makes the quantification of search results using errors from known theories ambiguous, since there's often another nearby theory.
Hence, in this appendix, we explain how we perform error analysis and quantify how good the algorithm, CMA-ES, performs.

To calculate the error of a search result (a point in space), a straightforward way is to simply calculate the Euclidean distances between the result and all known theories, and then select the shortest one as the final prediction.
However, we know that $h_{ext}$ defines the boundary condition of a process and the central charge $c$ classifies CFTs, which make them different roles from $\vec{h}_{int}$.
Hence, in our procedure, we first make sure the search result has similar ${h}_{ext}$ and $c$, and then calculate error in the full space.
With this in mind, the following is the detail on how we calculate the error that qualifies our search results.

To determine which known minimal model is the closest to a given search result, we proceed as following.
First, we apply a filter based on the relative errors in $c$, $h_{ext}$: Only minimal models whose deviations in both quantities are less than 5\% remain as candidates, that is, 
\begin{align}
    \frac{||\vec{v}-\vec{v}_{mm}||_2}{||\vec{v}||_2}<5\%
\end{align}
where $\vec{v}=\left(h_{ext}, c\right)$, $mm$ means the analytic values of minimal models, and we use L2 norm for the calculation.

Among the remaining candidates, we then compute the relative error in the full $(\vec{h}_{int}, h_{ext}, c)$ space and select the one with smallest error as the closest minimal model.
More explicitly, the relative error is defined as
\begin{align}
    min_i\left\{\frac{||\vec{u}_i-\vec{u}_{mm}||_2}{||\vec{u}_i||_2}\right\}
\end{align}
where $\vec{u}=\left(\vec{h}_{int}, h_{ext}, c\right)$, and $i$ labels the remaining candidates.

\pagenumbering{alph}

\twocolumngrid

\bibliography{refs}

\end{document}